\documentclass[sigconf, nonacm]{acmart}
\AtBeginDocument{%
  }

\usepackage{graphicx} % Required for inserting images
\usepackage{listings}
\usepackage[T1]{fontenc}        % better typefaces/language support
\usepackage{listings}           % core listing package
\usepackage{xcolor}             % colours for syntax + frame
\usepackage{algorithm}
\usepackage{algpseudocode}
\usepackage{amsmath}
\usepackage{url}
\usepackage{minted}
\usepackage{xcolor} 
\usepackage{hyperref}

\setminted{
    linenos=false,
    fontsize=\small,
    autogobble=true,
    style=sas,
}

\begin{document}

\title{Implicit Modeling for 3D-printed Multi-material Computational Object Design via Python}

\author{Charles Wade}
\email{charles.wade@colorado.edu}
\orcid{0000-0002-6056-7717}
\affiliation{%
  \institution{University of Colorado Boulder}
  \city{Boulder}
  \state{Colorado}
  \country{USA}
}
\affiliation{%
  \institution{Draper Scholars, The Charles Stark Draper Laboratory, Inc}
  \city{Cambridge}
  \state{Massachusetts}
  \country{USA}
}

\author{Devon Beck}
\email{dbeck@draper.com}
\orcid{0000-0001-5010-579X}
\affiliation{%
  \institution{The Charles Stark Draper Laboratory, Inc}
  \city{Cambridge}
  \state{Massachusetts}
  \country{USA}
}

\author{Robert MacCurdy}
\email{maccurdy@colorado.edu}
\orcid{0000-0002-1726-151X}
\affiliation{%
  \institution{University of Colorado Boulder}
  \city{Boulder}
  \state{Colorado}
  \country{USA}
}
\authornote{Corresponding Author}

\renewcommand{\shortauthors}{Wade et al.}

\begin{abstract}
This paper introduces open-source contributions designed to accelerate research in volumetric multi-material additive manufacturing and metamaterial design. We present a flexible Python-based API facilitating parametric expression of multi-material gradients, integration with external libraries, multi-material lattice structure design, and interoperability with finite element modeling. Novel implicit multi-material modeling techniques enable detailed spatial grading at multiple scales within lattice structures. Additionally, our framework integrates with finite element analysis, offering predictive simulations via adaptive mesh sizing and direct import of simulation results to guide material distributions. Practical case studies illustrate the utility of these contributions, including functionally graded lattices, algorithmically generated structures, and simulation-informed designs, exemplified by a multi-material bicycle seat optimized for mechanical performance and rider comfort. Finally, we introduce a mesh export strategy compatible with standard slicing software, significantly broadening the accessibility and adoption of functionality graded computational design methodologies for multi-material fabrication.
\end{abstract}




\begin{teaserfigure}
  \includegraphics[width=\textwidth]{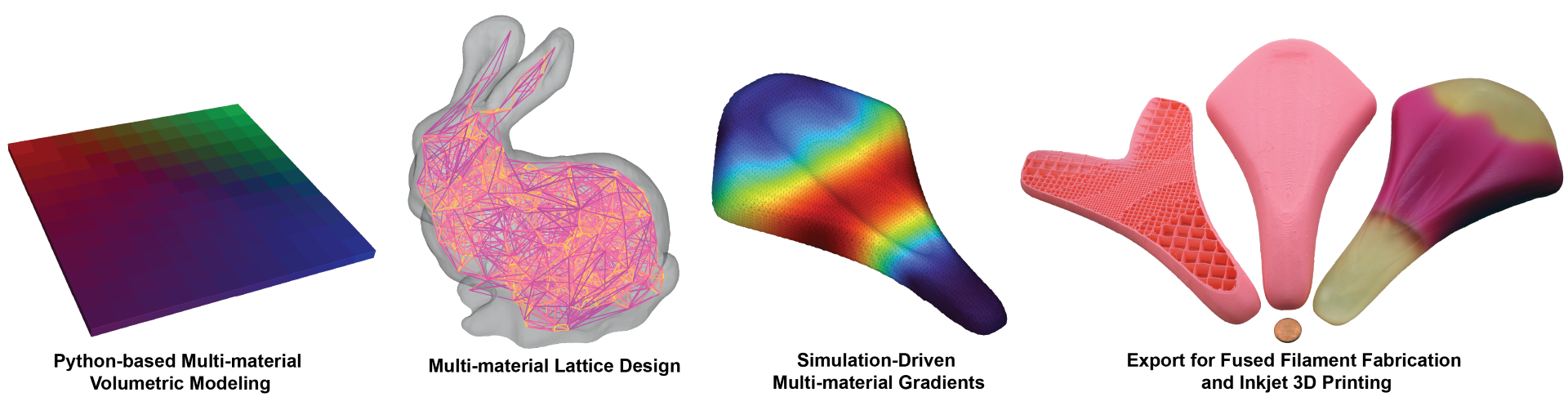}
  \label{fig:teaser}
\end{teaserfigure}

% \received{17 July 2025}

\maketitle

\section{Introduction}
The advancement of additive manufacturing (AM) technologies has enabled the creation of geometries and material combinations previously considered impractical or impossible. Particularly, multi-material AM systems now offer high-resolution control of spatially varying materials throughout printed objects. Despite these hardware advancements, computational design tools for AM have not adequately evolved to harness these capabilities. Existing CAD methodologies predominantly use boundary surface representations. This approach is suited best for homogeneous or discrete transitions between materials, limiting design complexity. Consequently, engineers and designers were constrained by CAD tools that do not readily support precise, volumetric distributions of multiple materials within complex geometries.

Volumetric multi-material design promises significant benefits across a wide range of applications, including compliant mechanisms and metamaterials. Unlike traditional homogeneous or discretely heterogeneous designs, functionally graded materials (FGMs) exhibit continuous spatial variations in composition, enabling tailored mechanical properties and performance optimization. The capability to design with continuously varying properties is increasingly critical for applications that demand spatially precise performance tuning, such as biomedical implants, customized ergonomics, and structural optimization.

OpenVCAD, an open-source volumetric design compiler, addresses these design limitations by offering implicit representations capable of expressing both geometry and spatially varying material distributions. However,  OpenVCAD exhibits several key limitations: it is targeted at primarily Inkjet-based AM systems, relies on a custom scripting language with limited expressiveness, and lacks integration with simulation methods for performance prediction. These limitations restrict broader adoption and hindered practical applications.

In this paper, we present novel methodologies as open source contributions to  the OpenVCAD framework. These methodologies significantly expand capabilities and applicability, addressing the shortcomings inherent in the current implementation. First, we introduce a substantial shift from a domain-specific scripting language to a flexible Python-based API. This transition facilitates greater parametric complexity, enabling designers to incorporate variables, loops, conditional logic, and external libraries, thus significantly enhancing design expressiveness and versatility.

Additionally, we introduce novel methods enabling the implicit design of multi-material lattice structures with spatially varying material properties. We demonstrate the efficacy of this approach through printed artifacts integrating compliant (soft) and stiff (rigid) materials within single lattice structures. Extending beyond static design, we provide a novel workflow for exporting OpenVCAD models for finite element analysis (FEA) simulation in ABAQUS, allowing prediction and validation of mechanical performance prior to fabrication. Through side-by-side comparisons between simulated and physical lattice structures, we demonstrate practicality of our approach.

Furthermore, we propose a novel methodology for importing FEA simulation results back into OpenVCAD, facilitating automated material grading based on simulation results. We illustrate this capability using a practical application: a functionally graded 3D-printed bicycle seat. By mapping simulated deformation from rider loads directly into spatial gradients of soft and rigid materials, we demonstrate a streamlined workflow from simulation-informed design to fabrication with targeted mechanical responses. This approach enables precise tailoring of compliance and performance characteristics.

Lastly, we introduce a mesh-based export strategy compatible with standard slicing software such as PrusaSlicer \cite{PrusaSlicer}. This method allows OpenVCAD designs, previously confined largely to Inkjet systems, to be realized using multi-material material extrusion workflows. Using our bicycle seat design, we demonstrate gradient-derived assignment of variable infill densities, optimizing mechanical performance to mitigate deformation.

Overall, these advancements collectively represent significant steps toward an integrated, simulation-informed, and broadly applicable design framework for volumetric multi-material additive manufacturing. Through enhanced scripting capabilities, comprehensive simulation integration, and expanded fabrication compatibility, we provide a robust and flexible method for researchers to use to investigate advanced multi-material 3D printing applications. 

\section{Related Work}
While recent advancements in multi-material additive manufacturing have enabled increasingly complex fabrication capabilities, design tools have struggled to keep pace. A wide range of design frameworks have been proposed to support multi-material material modeling, each offering different trade-offs in representation fidelity, scalability, and compatibility with manufacturing workflows. In this section, we review prior work across five main categories. We begin with homogeneous multi-material design methods, which model discrete material regions using boundary representations. We then examine voxel-based approaches that offer direct volumetric material control, followed by non-voxel volumetric methods that aim to improve efficiency and expressiveness. We next summarize the contributions and limitations of OpenVCAD, an implicit modeling framework tailored for multi-material design. Finally, we conclude with a discussion of programming-based CAD systems that motivate our move toward a general-purpose, scriptable design interface.

\subsection{Homogeneous Multi-material Design}
Historically, the most prevalent approach for multi-material additive manufacturing involves representing designs as multiple discrete solids, each with homogeneous material compositions assigned individually. This methodology aligns closely with conventional geometric representations employed in widely-used CAD systems such as SolidWorks and Fusion360, which used constructive solid geometry (CSG) and boundary representation (b-rep) to define and visualize objects \cite{SolidWorks, Fusion360}. Boundary representations partition surfaces into cells composed of vertices, edges, and faces, encapsulating both geometric and topological details \cite{Lienhardt}. Similarly, CSG methods describe solids through Boolean combinations of primitive shapes defined primarily by their surfaces and were originally developed for rendering applications \cite{csghistory}.

Despite their effectiveness in visualizing discrete, solid models, homogeneous boundary-based representations struggle significantly when expressing functionally graded materials. Representing continuously varying material distributions with a boundary representation requires segmentation into distinct volumes, each with independent boundary definitions. When capturing material gradients, the complexity of managing numerous boundary surfaces escalates rapidly, increasing computational demands for slicing and toolpath generation. Moreover, threshold-based methods inherently introduce abrupt material transitions, potentially creating structural weaknesses at these interfaces \cite{hasanov2020mechanical}. Thus, traditional homogeneous multi-material design approaches are fundamentally limited for advanced additive manufacturing processes requiring precise volumetric material control.

\subsection{Voxel-Based Methods}
Voxel-based representations provide an intuitive and direct method for capturing heterogeneous and graded material designs, commonly used in graphics and rendering applications such as medical imaging, fluid simulation, and volumetric rendering of phenomena like clouds or fire \cite{openVDB}. Typically, these methods store volumetric data on material concentration, opacity, or density within discretized voxel grids, which can be efficiently visualized using texture slicing or ray-tracing techniques \cite{adv_graphics_textbook}. However, these standard voxel representations face significant computational challenges in additive manufacturing contexts, particularly for high-resolution inkjet and polyjet 3D printing systems, whose build volumes may consist of billions of voxels.

To mitigate computational limitations, Museth introduced OpenVDB, a sparse voxel data structure aimed at significantly reducing memory footprints for volumetric data by efficiently storing sparse regions \cite{openVDB}. OpenVDB offers various tools for voxel-based modeling, including Boolean operations and affine transformations. Building on this, NanoVDB provides a GPU-native implementation, enhancing performance through parallel processing capabilities \cite{nanovdb}. Despite these advancements, both OpenVDB and NanoVDB are primarily tailored for rendering applications. They lack specialized functionalities needed to convert sparse voxel structures into manufacturable formats suitable for additive manufacturing processes. Additionally, as data structures rather than complete design frameworks, they lack user-friendly interfaces for engineers aiming to utilize voxel-based design methods directly. PicoGK, an extension built upon OpenVDB, provides a more complete workflow tailored specifically for engineering design using voxels \cite{Kayser2024PicoGK}. However, despite employing voxel-based representations, PicoGK does not support multi-material heterogeneous design, limiting its capability to represent functional gradients effectively.

Similarly, Brauer and Aukes developed VoxelFuse, a voxel-based CAD framework for multi-material additive manufacturing \cite{VoxelFuse, Brauer2019}. VoxelFuse leverages voxels to simultaneously define geometric and material information, offering capabilities for voxelization, simulation, and integration with common CAD software like OpenSCAD, SolidWorks, and Fusion360. The simulation modules within VoxelFuse rely on voxel-specific frameworks such as VoxCAD and Voxelyze to evaluate designs prior to manufacturing \cite{VoxCAD}. Nonetheless, VoxelFuse shares similar scalability limitations inherent to voxel-grid representations, as operations scale cubically with grid dimensions. Consequently, it becomes computationally prohibitive for large-volume and high-resolution printing processes.

\subsection{Non-voxel Methods}
In contrast to voxel-based representations, non-voxel approaches represent heterogeneous material distributions without discretizing the design domain, thereby improving computational efficiency. However, these methods often lack the robust tooling, accessibility, or direct applicability necessary for additive manufacturing.

Foundry and OpenFab exemplify non-voxel approaches for multi-material design, emphasizing volumetric texture rendering and color gradients \cite{OpenFab, Foundry}. OpenFab uses a two-phase pipeline, beginning with boundary surface definitions and subsequently applying volumetric textures and procedurally defined "fablets" to create detailed material distributions. Foundry similarly supports detailed texture synthesis and the creation of alloyed or functionally graded materials, but relies on externally defined geometries, limiting its integration with existing additive manufacturing workflows.

GraMMaCAD provides another approach, enabling interactive definition of material gradients on imported boundary-surface geometries \cite{grammacad2, grammacad}. While intuitive, GraMMaCAD's reliance on manual region selection limits its scalability when designs contain numerous complex sub-regions, as commonly found in advanced applications such as meta-materials. Thus, methods offering native volumetric representations paired with geometry definition in a programmatic way are preferred.

Elber et al. proposed using volumetric representations (V-reps) based on B-splines, which naturally facilitate porous and heterogeneous designs \cite{splines}. V-reps closely integrate geometric modeling with finite element analysis, supporting advanced applications in manufacturing functionally graded materials. Their methodology is implemented within the IRIT Modeling Environment, providing a specialized toolset for volumetric modeling.

Pasko et al.'s \textit{Constructive hypervolume modeling}, uses scalar fields combined with geometric primitives within a hierarchical constructive tree to efficiently represent heterogeneous materials \cite{hypervolume}. Pasko et al.'s approach closely resembles implicit modeling frameworks like BlobTrees, organizing primitives and material attributes hierarchically \cite{hypervolume_cont, blobtree0, blobtree1, blobtree2}. While theoretically powerful for rendering and volumetric texture definition, Pasko et al.'s methodology lacks explicit procedures for translating implicit representations into discretized formats required for additive manufacturing. Additionally, it does not provide a direct method for representing objects as volume fractions, limiting its immediate applicability in multi-material 3D printing.

\subsection{OpenVCAD: Implicit Representation for Multi-Material AM}
Wade et al. (2024) proposed OpenVCAD, an efficient implicit representation specifically designed for expressing complex, multi-material, and functionally graded objects in additive manufacturing~\cite{wade_openvcad_2024}. Presented for inkjet-based AM systems, OpenVCAD addresses the inherent limitations of conventional boundary-surface design methods, such as STL or 3MF files, which inadequately represent volumetric material distributions due to their reliance solely on discrete boundary surfaces. OpenVCAD employs an implicit approach for both geometry and multi-material composition. It defines objects using two functions: a signed distance function (geometry), and a collection of volume fraction functions (multi-material distribution). OpenVCAD employs a tree-like structure of nodes and operators that allow for importing or creating different geometries, performing CSG-like operations such as boolean combinations, transformation, and multi-material functions such as blending or grading. The implicit approach inherently supports infinite resolution scaling and affine transformations, enabling detailed representation of both geometry and spatially varying material gradients without loss of precision.

Wade et al. (2025) proposed a subsequent extension of OpenVCAD that demonstrated its applicability beyond voxel-based inkjet systems, introducing gradient-aware slicing methods explicitly targeting toolpath-based AM systems~\cite{wade_implicit_2025}. Their work showed how implicit material distributions could inform toolpath planning for multi-material and functionally graded prints, including scenarios involving filament mixing hotends and temperature-responsive foaming materials. This expansion highlighted OpenVCAD’s versatility but was limited to a narrow selection of printing processes supported by a custom slicer.

Despite these advancements, OpenVCAD presents several notable limitations. The framework uses a custom domain-specific language for defining objects, limiting designers due to the absence of conventional programming constructs such as variables, loops, and conditional statements. Likewise, although the Wade el al. (2024) mentioned potential integration with finite element analysis (FEA) simulations, comprehensive methods for exporting OpenVCAD-defined objects to simulation-ready meshes and subsequently re-importing FEA results are not robustly prescribed. Finally, the implementation of lattice modeling within OpenVCAD is limited to importing externally generated lattice meshes and applying functional grading after. Our work expands by proposing an implicit lattice design method that supports multi-material functional grading within structures. This work will build on the OpenVCAD framework, by presenting novel open-source contributors that address these limitations. 

\subsection{Programming Based Computer Aided Design}
Programming-based CAD methodologies have emerged as powerful tools for parametric design, overcoming limitations inherent to graphical user interface-driven CAD systems. In contrast to interactive methods that often restrict parametric complexity, programming-based CAD allows designers to define objects via scripts written in specialized or general-purpose languages.

One of the earliest examples of such a system is Hyperfun, introduced by Pasko et al.~\cite{hyperfun}, employing Functional Representation (F-rep) coupled with a specialized programming language. Hyperfun allowed designers to script parametric objects using geometric and mathematical functions. Extensions of Hyperfun addressed heterogeneous material distribution and micro-structure design for additive manufacturing~\cite{hypervolume, hypervolume_cont, paskolattice}. However, these extensions remained primarily theoretical or focused narrowly on rendering applications rather than additive manufacturing workflows. Another widely used programming-based CAD tool is OpenSCAD, which provides a domain-specific scripting language to construct solid geometries from boundary representations through Constructive Solid Geometry (CSG). In OpenSCAD, objects are defined by scripts, processed as CSG trees composed of primitive shapes like cubes and spheres combined through Boolean operations. Despite its robustness for single-material designs, OpenSCAD is constrained by underlying boundary-surface representations, making it impractical for defining intricate multi-material or functionally graded designs due to prohibitively large file sizes and compilation times. Both Hyperfun and OpenSCAD employ their own domain specific programming languages that are not directly compatible with each other, or with external tools. This limits their applicability, especially with complex computational design workflows that are data driven or algorithmic.  

SolidPython serves as a Python-based wrapper for OpenSCAD that provides users with improved compatibility and interoperability with external Python libraries \cite{solidpython}. By leveraging the extensive ecosystem of scientific computing tools available in Python, SolidPython enhances parametric modeling capabilities significantly beyond those achievable with OpenSCAD's  domain-specific language alone. However, is limited by the homogeneous and surface based modeling employed by the OpenSCAD kernel. Therefore it would be advantageous to have a CAD modeling framework that offers both multi-material heterogeneous design capabilities paired with an interface in a general purpose programming language such as Python, which we introduce here.

Another widely used parametric design environment is Rhinoceros 3D (Rhino) combined with Grasshopper \cite{McNeel_Rhino3D}. Rhino provides the underlying geometry kernel, while Grasshopper adds a visual, programming-based interface for parametric design \cite{Grasshopper3D_Website}. Grasshopper also supports embedding Python scripts within its node-based workflows, enabling interoperability with advanced Python logic \cite{IranNejad2024GHpython}. However, these capabilities remain constrained within the proprietary Rhino/Grasshopper ecosystem, limiting accessibility in open-source workflows. A range of plugins extend their functionality—for example, Crystallon for lattice design and Silkworm for G-code generation \cite{Porterfield_Crystallon_About, ProjectSilkworm_Silkworm}. Despite their versatility, Rhino, Grasshopper, and associated plugins rely on boundary surface representations and do not support volumetric multi-material gradients, restricting their applicability to multi-material 3D printing and design.

\section{Methods and Case Studies}
The Methods and Case Studies section presents our approach to programmable multi-material additive manufacturing. First, we detail how transitioning from a custom domain-specific language (DSL) to Python enables dynamic and expressive multi-material design capabilities, supporting parametric, procedural, and algorithmic workflows. This section includes illustrative examples demonstrating the integration of Python libraries for complex design patterns. Subsequently, we explore advanced multi-material lattice generation methods, introducing novel primitives and hierarchical gradient strategies for designing spatially graded lattices. Next, we describe our methodology for exporting volumetric designs to finite element analysis workflows, highlighting techniques for adaptive meshing and simulation-informed multi-material design. Next, we introduce a method for importing  results directly into volumetric multi-material designs informed by predictive simulations. We conclude by introducing a new mesh export module that segments functionally graded designs into discrete triangulated meshes, enabling compatibility with existing slicer workflows. Each subsection includes targeted case studies that exemplify practical applications and validate the efficacy of the described methods.

\subsection{Programming Multi-Material Designs in Python}
OpenVCAD was initially implemented around a custom domain-specific language (DSL) designed to describe multi-material implicit geometries. This early language focused purely on structural geometric description and lacked the expressiveness of a general-purpose language. Notably, it did not support basic programming constructs such as variables, loops, or conditionals. As a result, it was difficult to create parametric or algorithmic designs within the language itself. Designers are forced to use another language, such as Python or Matlab, to generate OpenVCAD code indirectly as text files, resulting in verbose workflows. The DSL lacked an interchange standard, preventing interaction with external tooling, data sources, or libraries, limiting its potential in automated design workflows.

To address these limitations, we introduce \texttt{pyvcad}, a Python library that enables dynamic and programmable construction of OpenVCAD design trees. \texttt{pyvcad} provides a Python interface for the volumetric design, rendering, and 3D printer compilers outlined in existing OpenVCAD works \cite{wade_openvcad_2024, wade_implicit_2025}. Implemented as bindings to the C++ based OpenVCAD library, \texttt{pyvcad} provides a balance of efficiency and speed matched with an easy to use Python interface. The library enables users to construct design graphs using standard Python syntax and control flow, allowing for procedural and data-driven design workflows. Crucially, this opens the door to integration with scientific Python libraries such as NumPy, SciPy, or Pandas, allowing OpenVCAD models to be informed by datasets, simulation outputs, or optimization results.

While CAD libraries such as SolidPython, Trimesh, and pythonOCC support parametric geometric modeling within Python, none of these tools offer volumetric representations with continuous, multi-material gradients. Moreover, they rely primarily on surface modeling paradigms and lack an implicit representation for both geometry and material distribution. Commercial tools such as nTopology provide implicit geometry modeling but do not support implicit multi-material definitions or open, programmable Python APIs \cite{nTop}. In contrast, \texttt{pyvcad} is an open-source framework to support fully parametric, implicit, and functionally graded multi-material volumetric design in Python. This Python interface for OpenVCAD enables a broad set of new workflows, documented in the case studies, including procedural spatially varying lattice generation, stress-informed material mappings, and integration with external libraries and algorithms.

\subsubsection{Case Studies: Programmed Multi-material Design}
We present several example scripts that demonstrate the expressiveness and flexibility of \texttt{pyvcad}. These case studies show how users can define parametric models using standard programming techniques while taking full advantage of OpenVCAD's volumetric, multi-material design infrastructure.

\begin{figure}[h]
    \centering
    \includegraphics[width=\linewidth]{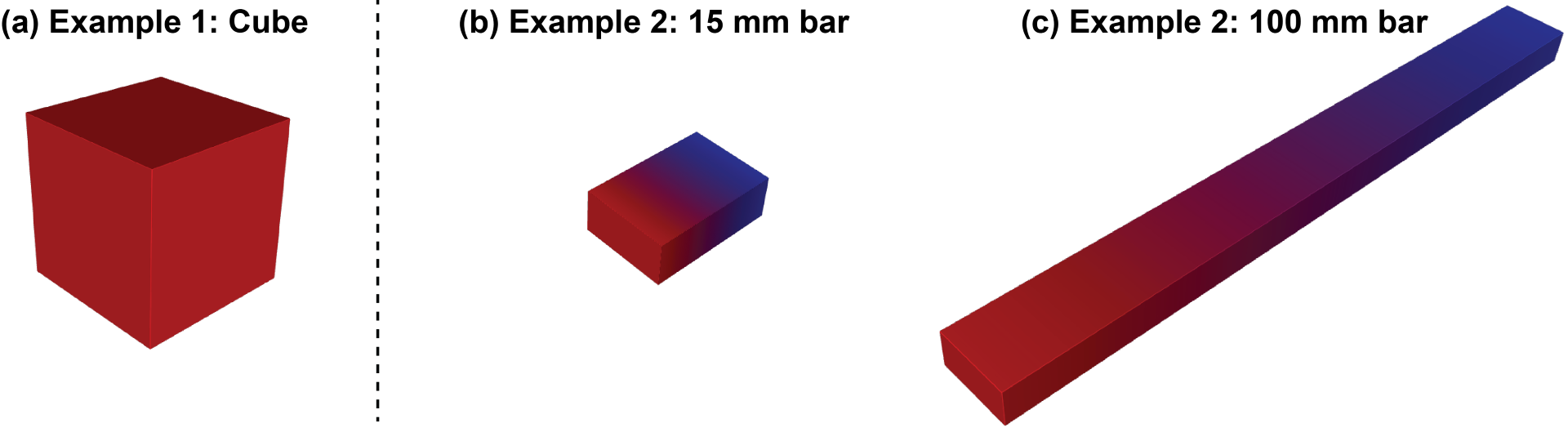}
    \caption{(a) Simple single material cube geometry created from the script given in listing \ref{lst:rect_prism_code} (b) 15\,mm and (c) 100\,mm bar with a red-to-blue gradient created using the script given in listing \ref{lst:graded_prism_code}.}
    \label{fig:graded-bars}
\end{figure}

Code Listing \ref{lst:rect_prism_code} and Figure \ref{fig:graded-bars}(a) show an example that is a minimal "Hello World" script that generates a red rectangular prism centered at the origin. The geometry is defined parametrically using Python variables, and the visualization module \texttt{pyvcadviz} is used to render an interactive preview of the resulting volume. This example demonstrates the simplicity of constructing design graphs using parameterized inputs.

\begin{listing}[h]
    \caption{Example \texttt{pyvcad} Code for Rectangular Prism}
    \label{lst:rect_prism_code}
    \begin{minted}{python}
    import pyvcad as pv
    materials = pv.default_materials()
    
    # Parameters
    center_point = pv.Vec3(0, 0, 0) # At origin
    dimensions = pv.Vec3(10,10,10)  # In mm
    # Create a Rectangular Prism
    root = pv.RectPrism(center_point, dimensions, 
                        materials.id("red"))
    \end{minted}
\end{listing}

The second example, shown in Code Listing \ref{lst:graded_prism_code} and Figure \ref{fig:graded-bars}(b-c) introduces a functionally graded design, wherein a bar transitions from red to blue along its length. This is achieved using the \texttt{FGrade} node, which accepts a list of expressions representing the volume fractions of materials, the corresponding materials, a mode selection flag (probabilistic mixing or thresholding), and the geometry to which the gradient is applied. In this case, a gray rectangular prism is created and then overridden with a red-to-blue gradient that varies continuously along the $x$-axis.

To ensure the design scales appropriately with changes in geometry size, the volume fraction expressions are defined as parameterized math strings using Python f-strings. This approach ensures that gradient behavior remains consistent even if the length of the bar is modified. While we support passing Python function pointers instead of strings, we strongly encourage the use of math strings, which are compiled into efficient abstract syntax trees using the \texttt{exprtk} library \cite{Partow}. This results in significantly better runtime performance.

\begin{listing}[h]
    \caption{Example \texttt{pyvcad} Code for Graded Bar}
    \label{lst:graded_prism_code}
    \begin{minted}{python}
import pyvcad as pv
materials = pv.default_materials()

# Parameters
dimensions = pv.Vec3(15,10,5) # in mm
center = pv.Vec3(0,0,0)
# Create a gray bar
bar = pv.RectPrism(center, dimensions, materials.id("gray"))
# Apply a red to blue gradient
expressions = [f"x/{dimensions.x}+0.5", 
               f"-x/{dimensions.x}+0.5"]
materials = [materials.id("red"), 
             materials.id("blue")]
root = pv.FGrade(expressions, materials, True, bar)
    \end{minted}
\end{listing}

The final example demonstrates a more complex use case involving a parametric color calibration sheet for inkjet 3D printers. The code for this example is given in listing \ref{lst:swtaches_code}. The goal of this design is to create a sheet composed of spatially varying mixtures of three base materials. Each swatch in the grid corresponds to a unique volume fraction of the three base materials, providing a known color reference for printer calibration. The design is defined by a reusable Python function that accepts parameters such as swatch size, number of swatches in each dimension, thickness, and material palette. Inside the function, a nested loop iterates over the grid and assigns the appropriate volume fractions to each swatch using the \texttt{FGrade} node. The resulting child nodes are then aggregated into a single design using a \texttt{Union} node.

\begin{figure}[h]
    \centering
    \includegraphics[width=\linewidth]{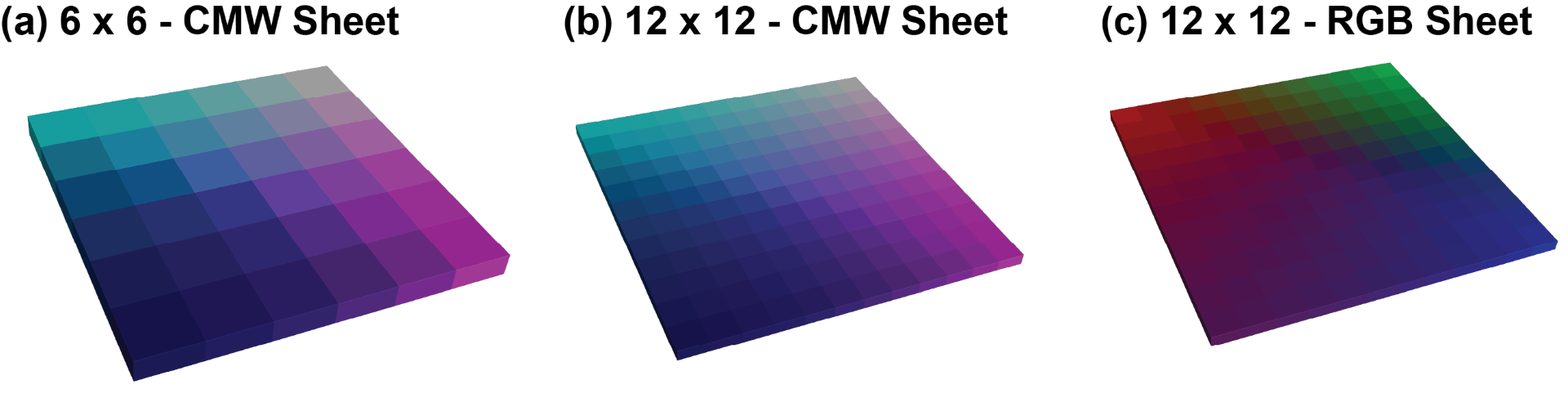}
    \caption{Color calibration sheets created using \texttt{pyvcad}. (a) 6$\times$6 grid using cyan, magenta, and white (CMW). (b) 12$\times$12 CMW sheet. (c) 12$\times$12 grid with red, green, and blue.}
    \label{fig:color-sheets}
\end{figure}

This example highlights several advanced features enabled by \texttt{pyvcad}, including user-defined functions for encapsulating design logic, dynamic control flow for scalable pattern generation, and the ability to combine large numbers of primitives into a single structure. Similarly, it demonstrates the multi-material parametric functionally of \texttt{pyvcad}. Figures \ref{fig:color-sheets}a--c show calibration sheets generated using different parameters, demonstrating the flexibility and reusability of the parametric function.

\begin{listing}[h]
    \caption{\texttt{pyvcad} Code for Parametric Color Swatches}
    \label{lst:swtaches_code}
    \begin{minted}{python}
import pyvcad as pv
def create_calibration_sheet(s, count_x, count_y, 
                             thickness, materials):
    # center the whole sheet around the origin
    w  = s * count_x
    h = s * count_y
    union = pv.Union()
    for i in range(count_x):
        for j in range(count_y):
            # compute volume fractions
            c_frac = i / (count_x - 1)
            m_frac = j / (count_y - 1)
            w_frac = 1.0 - (c_frac + m_frac)
            fractions = [f"{c_frac:.3f}",
                         f"{m_frac:.3f}",
                         f"{w_frac:.3f}"]
            # compute the center of this swatch
            x_pos = -w/2 + s/2 + i * s
            y_pos = -h/2 + s/2 + j * s
            center = pv.Vec3(x_pos, y_pos, 0)

            size = pv.Vec3(s, s, thickness)
            base = pv.RectPrism(center, size)
            graded = pv.FGrade(fractions, materials, 
                               True, base)
            union.add_child(graded)
    return union

# Load some materials
mats = pv.default_materials()
root = create_calibration_sheet(
    s         = 25, # swatch size (mm)
    count_x   = 12, # subdivisions
    count_y   = 12, 
    thickness = 10, # mm plate thickness
    materials = [mats.id("cyan"), mats.id("magenta"),
                 mats.id("white")]
)
    \end{minted}
\end{listing}

\subsection{Multi-material Lattice Design}

While basic lattice structures have been demonstrated with OpenVCAD, robust methods for graph-based or strut-based lattices are notably absent \cite{wade_openvcad_2024}. We introduce enhanced lattice construction capabilities centered around a robust primitive for individual lattice elements: a cylindrical strut defined by two endpoints and capped at both ends. These struts form the basic building blocks of more complex lattice structures.

\begin{figure}[h]
    \centering
    \includegraphics[width=0.6\linewidth]{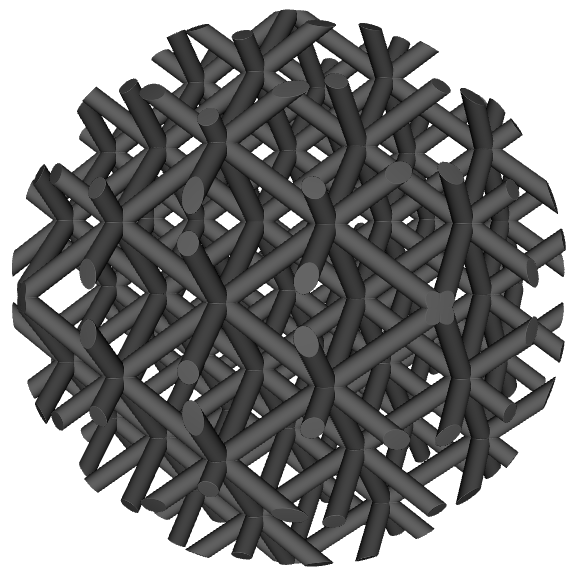}
    \caption{The result of code listing \ref{lst:tile_node_code} which tiles and intersections a body-centered cubic lattice withing a sphere.}
    \label{fig:lattice_sphere}
\end{figure}

To construct complex unit cells, we introduce the \texttt{GraphLattice} node, which accepts a list of vertex pairs to instantiate multiple struts simultaneously and merges them into a single geometry via a union operation. The \texttt{GraphLattice} node also supports a set of predefined lattice types, including Body-Centered Cubic (BCC), Face-Centered Cubic (FCC), and other standard topologies commonly used in additive manufacturing. An efficient design pattern involves constructing a single unit cell using \texttt{GraphLattice} and then replicating it periodically using the \texttt{Tile} node. The new \texttt{Tile} node replicates both the geometry and material distribution of its child node at a user-defined interval. To constrain the tiled lattice to a specific region, an \texttt{Intersection} node can be used to clip the repeated structure against a bounding volume. As shown in Listing~\ref{lst:tile_node_code} and Figure~\ref{fig:lattice_sphere}, this approach enables the construction of a tiled BCC lattice confined to a spherical domain.

Although it is possible to create an array of unit cells manually by translating and unioning individual instances, this approach is significantly less efficient. The \texttt{Union}-based method requires duplicating each instance in memory, resulting in both increased computational cost and memory usage. In contrast, the \texttt{Tile} node retains a single copy of its child, leading to substantial performance gains. However, a key limitation of the tiling approach is that it assumes uniformity across the geometry and material composition of the unit cell. In the following section, we describe strategies for introducing functional grading into lattice structures.

\begin{listing}[h]
    \caption{\texttt{pyvcad} Code Volume Filling Lattice}
    \label{lst:tile_node_code}
    \begin{minted}{python}
import pyvcad as pv
materials = pv.default_materials()

# Parameters
sphere_radius = 10
cell_size = pv.Vec3(5,5,5)
cell_type = pv.LatticeType.BodyCenteredCubic
strut_dia = 0.35

# Create a unit cell (BCC lattice)
cell_bcc = pv.GraphLattice(cell_type, cell_size, strut_dia,
                           materials.id("gray"))
# Tile the unit cell to fill space
lattice_fill = pv.Tile(cell_bcc) 
# Create bounding geometry (sphere)
sphere = pv.Sphere(pv.Vec3(0,0,0), sphere_radius)
# Intersection of lattice with bounding sphere
filled_sphere = pv.Intersection(False, [lattice_fill, sphere])
root = filled_sphere
    \end{minted}
\end{listing}

In multi-material design contexts, engineers and designers often seek to locally tune material properties such as stiffness and compliance to achieve specific mechanical responses. Our proposed lattice construction framework supports this capability by enabling functionally graded material distributions to be applied at multiple hierarchical levels: globally across the entire lattice, per individual unit cell, or at the level of each strut.

\begin{figure*}[h]
    \centering
    \includegraphics[width=\linewidth]{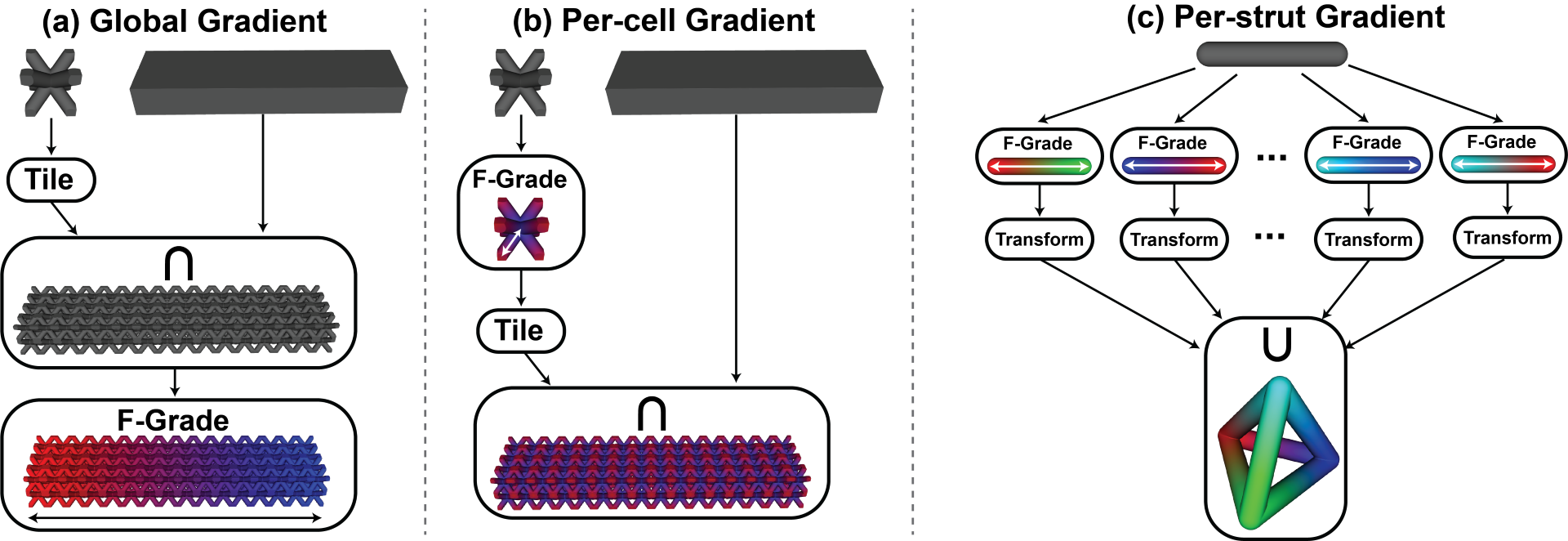}
    \caption{Illustrations of the three grading strategies supported by OpenVCAD: (a) global gradient, (b) local per-unit-cell gradient, and (c) local per-strut gradient.}
    \label{fig:gradient-types}
\end{figure*}

Figure~\ref{fig:gradient-types} depicts these three gradient strategies. The global gradient approach first constructs the entire lattice and subsequently applies a spatially varying gradient across the whole structure. In contrast, a local per-unit-cell gradient approach applies grading directly at the unit cell level, allowing each cell to repeat its internal gradient throughout the tiled structure. Finally, the per-strut grading method applies unique gradients to individual struts before combining them into a unit cell. This fine-grained control enables engineering of local structural performance that we will demonstrate in a future section on simulation.

\subsubsection{Case Study: Algorithmic Multi-material Lattice Generation}

To showcase the versatility and scalability of our enhanced lattice framework, we performed a complex case study involving the well-known Stanford Bunny model. This example demonstrates the integration of \texttt{pyvcad} with external Python libraries and algorithms to produce intricate, multi-material lattice structures within an arbitrary geometric domain. The algorithm used for generating this lattice is detailed in Algorithm~\ref{alg:lattice-bunny}.

\begin{algorithm}[h]
\caption{Algorithm for creating an algorithmically generated multi-material lattice}
\label{alg:lattice-bunny}
\begin{algorithmic}[1]
\State Import the Stanford Bunny as a mesh OpenVCAD node
\State Sample $N$ random points inside the mesh
\State Using \texttt{scipy}, compute the Delaunay triangulation of the random points
\State Extract a set of unique edges from the triangulation's simplices (tetrahedra)
\State Prune any edges that intersect or exit the bunny mesh
\State Determine the minimum and maximum edge lengths
\ForAll {edges}
    \State Create a strut node from edge endpoints
    \State Compute the magenta-to-yellow color gradient based on strut length:
    \Statex \quad shortest struts are pure yellow, longest are pure magenta, intermediate lengths proportionally graded
    \State Add graded strut to a union node
\EndFor
\State Take the union between the generated lattice union node and the bunny mesh node
\end{algorithmic}
\end{algorithm}

\begin{figure}[h]
    \centering
    \includegraphics[width=\linewidth]{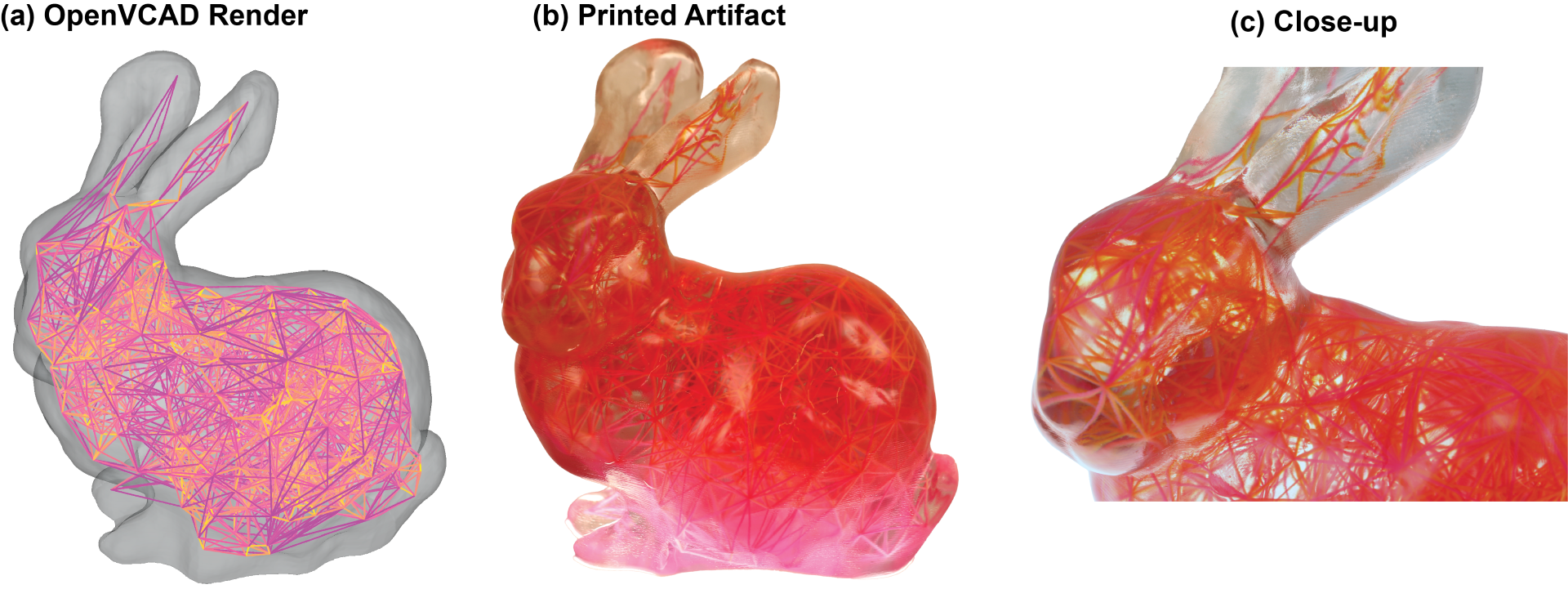}
    \caption{Algorithmic lattice generation case study: (a) OpenVCAD software render of the multi-material lattice-filled Stanford Bunny; (b) physical artifact printed on a Stratasys J750 printer; and (c) close-up view detailing graded internal lattice structures.}
    \label{fig:bunny-study}
\end{figure}

\begin{figure}[h]
    \centering
    \includegraphics[width=\linewidth]{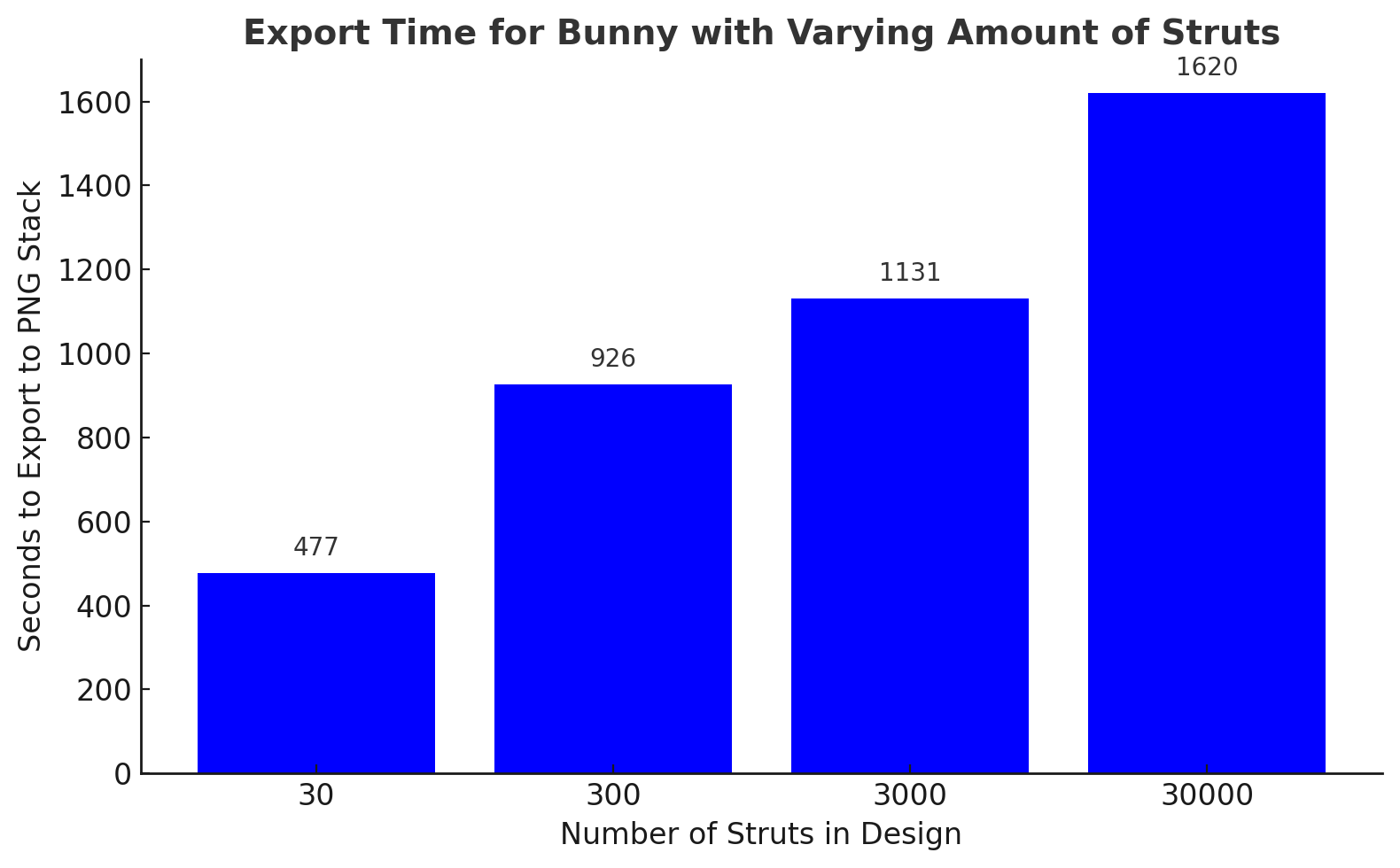}
    \caption{Export times for the bunny design with increasing numbers of struts. Each export sampled 10 billion voxels into PNG stacks. Despite a 1000-fold increase in strut count, export time increased by only 3.4x, highlighting the scalability of our functionally graded strut-design method for complex models.}
    \label{fig:bunny-performace}
\end{figure}

The resulting lattice structure comprises 3,289 uniquely graded struts, each individually colored according to length-based material blending between yellow and magenta. Figure~\ref{fig:bunny-study}(a) provides a rendering of the design generated in OpenVCAD, Figure~\ref{fig:bunny-study}(b) depicts the physical artifact fabricated on a Stratasys J750 printer using clear, magenta, and yellow materials, and Figure~\ref{fig:bunny-study}(c) highlights a detailed view of the bunny's facial region. The voxelization slicing step required sampling 10 billion voxels, completing in 19 minutes on a desktop equipped with a Ryzen 7 7700X processor. To further assess the performance of our multi-material, functionally graded strut-design method, Figure~\ref{fig:bunny-performace} illustrates the wall-clock time needed to slice the bunny model as the number of struts increases. Notably, despite a 1000-fold increase in strut count, export time increased by only a factor of 3.4. This analysis demonstrates that our lattice framework scales efficiently to support thousands of individually graded elements while highlighting the practical benefits of integrating OpenVCAD workflows with external Python libraries for advanced multi-material volumetric design.

\subsection{Export to Finite Element Analysis}

In multi-material additive manufacturing workflows, predicting the mechanical performance of functionally graded structures prior to fabrication is crucial. This is especially the case when the base materials exhibit significantly different  mechanical properties, such as soft elastomers versus rigid polymers. To facilitate performance prediction, we present a novel method to export functionally graded designs into finite element analysis (FEA) meshes compatible with commercial software such as Abaqus via \texttt{.INP} files. This exporter supports two types of FEA elements commonly used in solid mechanics simulations: brick elements (C3D8R) and tetrahedral elements (C3D4).

The FEA export process treats geometry and material assignment as distinct stages. First, we sample the implicit geometric representation of an OpenVCAD design to discretize the domain into finite elements. For brick elements, the domain is sampled on a structured grid within the bounding box of the OpenVCAD design. An element is created if its corresponding grid sample lies within or on the boundary of the implicit volume. Alternatively, for tetrahedral elements, we leverage the Computational Geometry Algorithms Library (CGAL) to generate a space filling tetrahedral mesh from the implicit function \cite{cgal_mesh3d, cgal_main}.

\begin{figure}[h]
    \centering
    \includegraphics[width=\linewidth]{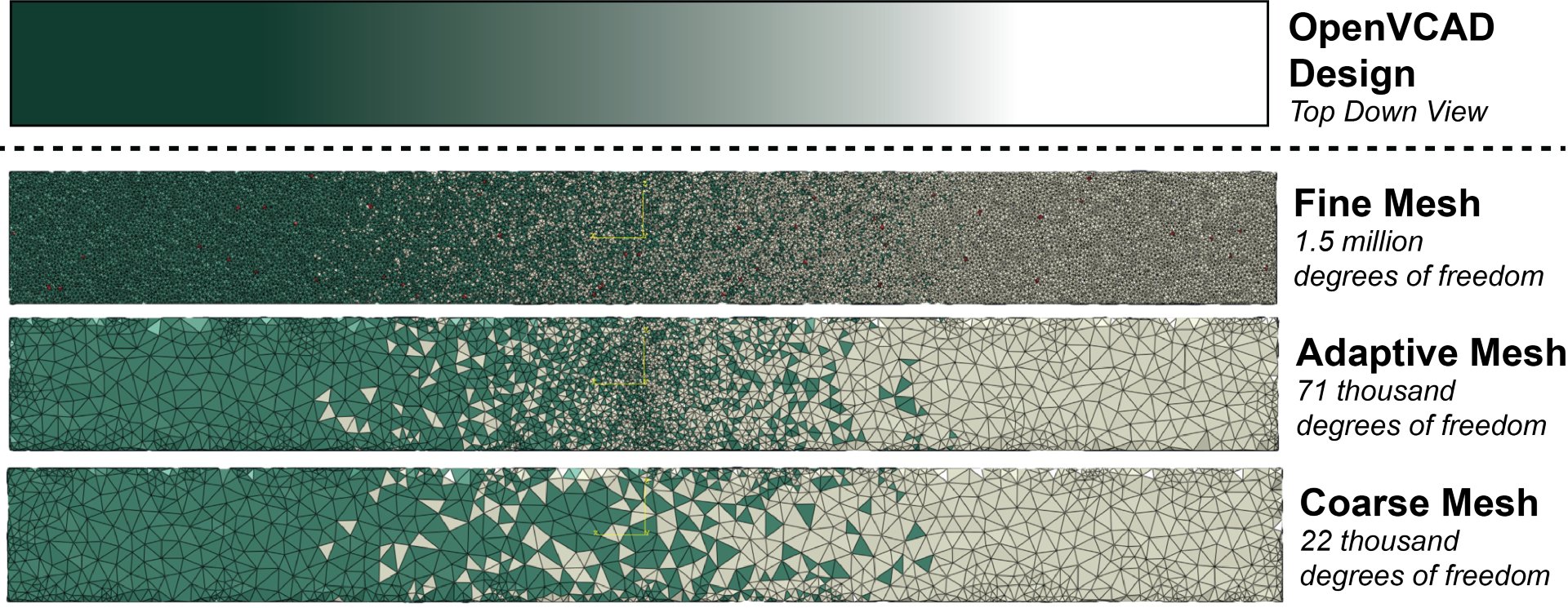}
    \caption{Comparison of finite element discretizations for a linear gradient bar using a coarse mesh (22 thousand DOFs), adaptive mesh (71 thousand DOFs), and fine mesh (1.5 million DOFs).}
    \label{fig:adaptive-fea}
\end{figure}

Following geometry discretization, each element (brick or tetrahedral) must be assigned a discrete material. We accomplish this by evaluating the multi-material distribution at the centroid of each element, producing a set of volume fractions. Following the probabilistic method described in the original OpenVCAD framework~\cite{wade_openvcad_2024}, these fractions serve as probabilities for assigning each material to an element. Figure~\ref{fig:adaptive-fea} shows a gradient interdigitated into discrete material assignments.  

We adopt this discretization approach because it reflects the current capabilities of OpenVCAD, which represents designs as combinations of base materials rather than as property fields. While interpolating material properties across volume fractions reduces mesh dependence, OpenVCAD does not directly model properties of mixtures. Instead, it exports voxelized, interdigitated designs that mirror the behavior of multi-material printing systems (e.g., inkjet), leaving it to the designer to experimentally characterize the resulting mixtures. Beginning with these low-level interdigitated models is essential for building a foundation of simulated and experimentally verified data, which can ultimately support higher-level models that predict the effective properties of material blends. Meisel et al. further showed that homogenized models fail to capture the true behavior of inkjet-printed multi-material objects once the microstructural features exceed 2 mm \cite{meisel2018impact}.

However, accurately capturing spatial gradients with this probabilistic method introduces a critical challenge: element size selection. A fine discretization with many small elements captures gradients accurately but dramatically increases computational demands, whereas coarser discretization sacrifices gradient fidelity. Figure~\ref{fig:adaptive-fea} illustrates this tradeoff using a simple linear gradient bar transitioning from green to white. To resolve this challenge, we introduce adaptive mesh sizing, which adjusts element size according to the local gradient complexity. Adaptive meshing is a feature that is commonly supported by FEA meshing software, but is usually employed to define element size to ensure geometry with fine detail is captured.

Our adaptive meshing approach uses a user-defined sizing field, expressed mathematically through variables and is provided provided to CGAL meshing routine. Specifically, the sizing field is defined as a function of the local material distribution heterogeneity, denoted by $h(x,y,z)$. The heterogeneity $h$ is computed at a point $(x,y,z)$ by evaluating the deviation of the local material fractions $f_i$ from a uniform distribution across $n$ materials:
\begin{equation}
h(x,y,z) = \frac{\sqrt{\sum_{i=1}^n \left(f_i - \frac{1}{n}\right)^2}}{\sqrt{1 - \frac{1}{n}}}
\end{equation}
Here, $h(x,y,z)$ ranges from 0 (homogeneous) to 1 (maximally heterogeneous). Users specify the minimum and maximum element sizes (\texttt{min\_cell} and \texttt{max\_cell}), and define the local cell length $L(x,y,z)$ as a function of $h$. Although a user can define complex non-linear math expressions, we will use the following simple linear mapping:
\begin{equation}
L(x,y,z) = \mathrm{min\_cell} + h(x,y,z)\,(\mathrm{max\_cell} - \mathrm{min\_cell})
\end{equation}
This adaptive sizing ensures finer discretization where there is a high mixture of elements, and larger elements when the design is more homogeneous. This provides a tunable balance between accuracy and computational efficiency (Figure~\ref{fig:adaptive-fea}).

\subsubsection{Case Study: Simulating Multi-material Lattices}

To demonstrate the  application of the OpenVCAD FEA exporter, we performed simulations on a body-centered cubic (BCC) lattice structure measuring $50\,\text{mm}\times50\,\text{mm}\times50\,\text{mm}$ composed of 5 unit cells per axis, each $10\,\text{mm}$ in dimension. The lattices were exported using the brick element exporter without adaptive meshing, resulting in uniform discretizations of approximately 240,000 elements per structure. A two material gradient with rigid and soft base materials was used. Five functionally graded lattices were studied: (1) homogeneous soft material, (2) a linear gradient transitioning from completely soft at the bottom to fully rigid at the top, (3) a Gaussian gradient peaking with soft material at the center, (4) a sinusoidal gradient introducing two soft regions, and (5) per-cell local radial gradients transitioning from rigid at the cell centers to soft at cell boundaries.

\begin{figure*}[h]
    \centering
    \includegraphics[width=\linewidth]{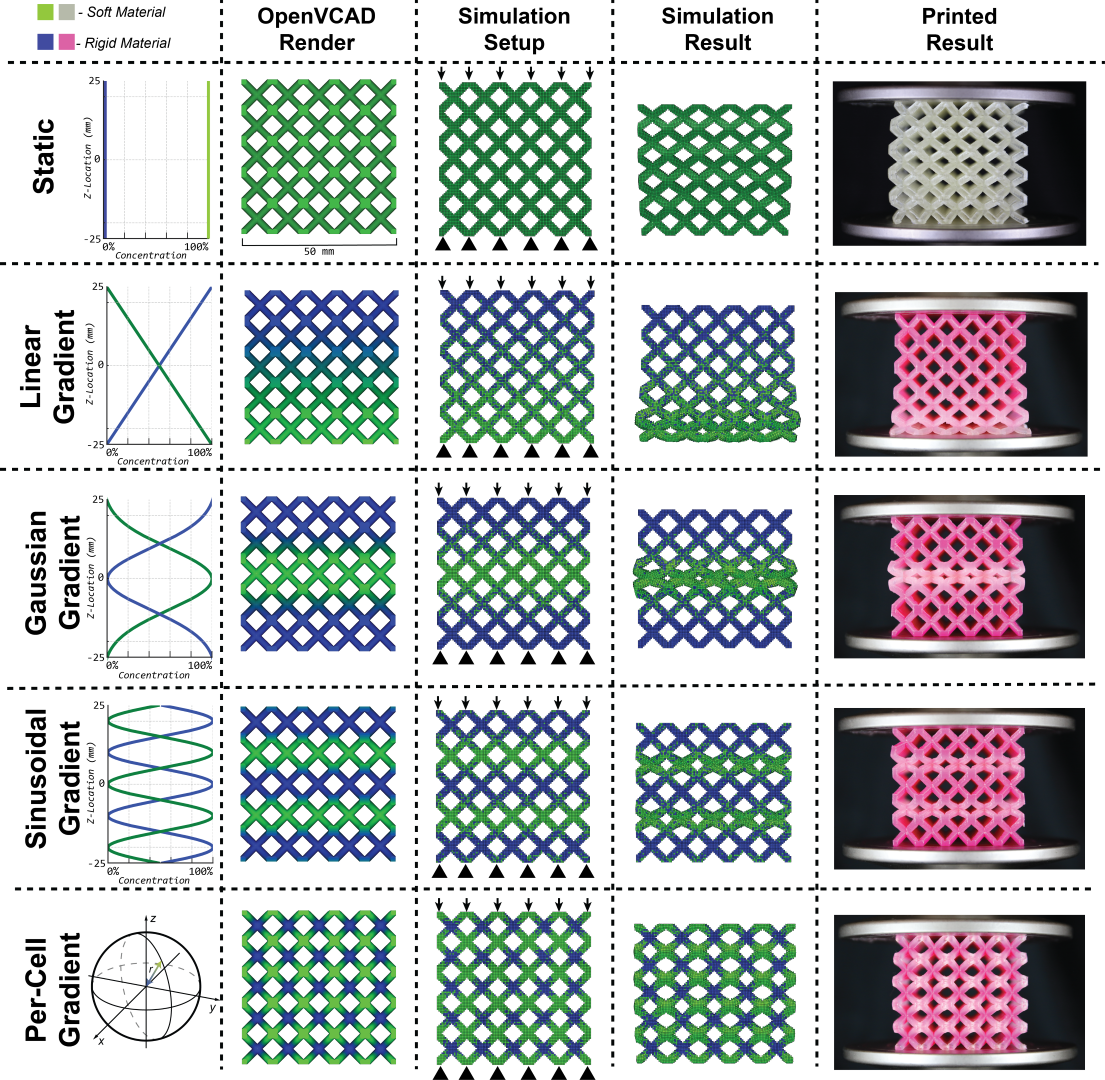}
    \caption{Case study of FEA simulation for multi-material lattices: (left-to-right) OpenVCAD design renders, corresponding FEA mesh with material assignment, Abaqus simulation results, and fabricated specimens printed with rigid (Vero) and soft (Agilus30) materials. Rows correspond to different gradient strategies as described in the text.}
    \label{fig:lattice-fea}
\end{figure*}

Figure~\ref{fig:lattice-fea} compares these lattices, showing OpenVCAD renders, corresponding FEA meshes with material sets (soft and rigid), Abaqus simulation results, and physical prints. Abaqus simulations utilized linear elastic material models (rigid material: $E=2850\,\text{MPa}$, $\nu=0.39$; soft material: $E=0.383\,\text{MPa}$, $\nu=0.50$). The values for these models were chosen based on the work of Majca-Nowak et al. for the rigid material and Qureshi et al. for the soft material \cite{majca-nowak_analysis_2023, qureshi_why_2022}. We used a static linear simulation. The boundary conditions consisted of fixing the bottom faces of the lattices while imposing a $5\,\text{mm}$ displacement at the top. As expected, simulation results show greater deformation in regions containing predominantly softer material, validated by physical specimens fabricated on a Stratasys J750 PolyJet printer using Vero (rigid) and Agilus30 (soft) materials.

This case study demonstrates OpenVCAD's capability to seamlessly integrate multi-material lattice design and predictive simulation, enabling informed design decisions prior to fabrication. We are not presenting an exhaustive study on high fidelity multi-material FEA simulation, but rather showing a demonstration of the framework to interface OpenVCAD with simulation software.

\subsection{Importing Simulation Results into OpenVCAD}

Simulation-driven design has become a cornerstone method for optimizing the structural performance of complex objects. Simulation is particularly valuable when integrating multi-material functionally via graded materials. While established generative design tools, such as nTopology, have successfully employed simulation results to guide geometric topology, they offer no support for simulation of multi-material gradients. Likewise, nTopology does not provide a method to map simulation results into multi-material gradients. To address this limitation, we introduce the novel \texttt{SimulationField} node for OpenVCAD, enabling direct import of finite element analysis (FEA) results and subsequent mapping into functionally graded multi-material distributions.

An OpenVCAD node must provide two essential outputs: (1) a geometric form represented by a signed distance function and (2) a multi-material volume fraction distribution. To efficiently meet these requirements, especially at resolutions required by high-resolution 3D printers that may demand billions of queries, the \texttt{SimulationField} node preprocesses simulation data into suitable intermediate structures.

The \texttt{SimulationField} node requires the user to specify the original FEA mesh, in this case a tetrahedral 3D mesh, used during simulation. Additionally, the user must provide a CSV file mapping each node in the tetrahedral mesh to corresponding simulation results, such as displacement values. This mapping enables the node to interpolate simulation data into spatially varying material gradients. The underlying FEA simulation can involve either single or multiple materials, as the \texttt{SimulationField} node translates physical simulation outputs (e.g., stress, displacement) into a new multi-material distribution.

To derive the signed distance form, we first extract a triangulated surface mesh from the imported tetrahedral FEA mesh by identifying faces that are not shared between adjacent tetrahedra. Subsequently, this surface mesh is converted into a sparse voxel grid using OpenVDB, encoding the signed distance to the object's surface. A user-selected coarse grid resolution (e.g., 0.1 mm) is chosen to balance memory usage and computational complexity. Signed distance queries at higher resolutions (e.g., 0.027 mm for PolyJet printers) employ efficient box interpolation within the sparse voxel representation to determine intermediate values.

For the material distribution form, we use a volumetric interpolation approach based on the simulation results provided at the nodes of the tetrahedral mesh. The algorithm for evaluating the material distribution at arbitrary points within the volume is summarized in Algorithm~\ref{alg:simfield-mat-dist}:

\begin{algorithm}[h]
\caption{Interpolating Simulation Results for Material Distribution}
\label{alg:simfield-mat-dist}
\begin{algorithmic}[1]
\Require Tetrahedral mesh $\mathcal{T}$, nodal simulation results $\mathbf{R}$, query point $\mathbf{q}$, material mapping expressions $\mathcal{E}$
\Ensure Multi-material distribution $M_{\mathbf{q}}$
\State Find tetrahedron $t \in \mathcal{T}$ containing point $\mathbf{q}$
\State Retrieve results $\mathbf{R}_i$ at vertices of $t$
\State Compute barycentric coordinates $(\lambda_1,\lambda_2,\lambda_3,\lambda_4)$ of $\mathbf{q}$ within $t$
\State Interpolate simulation result: $\mathbf{R}_{\mathbf{q}} \gets \sum_{i=1}^{4}\lambda_i\mathbf{R}_i$
\State Evaluate each expression in $\mathcal{E}$ using interpolated result $\mathbf{R}_{\mathbf{q}}$
\State Store volume fractions into distribution map $M_{\mathbf{q}}$
\State \Return $M_{\mathbf{q}}$
\end{algorithmic}
\end{algorithm}

A significant computational challenge arises from rapidly determining which tetrahedron contains a query point, especially for finite element meshes containing large numbers of elements. Naively checking every tetrahedron for containment would become prohibitively expensive, particularly with high element count FEA meshes. To improve performance, we employ an Axis-Aligned Bounding Box (AABB) tree, specifically utilizing the implementation provided by the Computational Geometry Algorithms Library \cite{cgal_aabb,cgal_main}. An AABB tree is a hierarchical spatial data structure that recursively partitions space into axis-aligned bounding boxes, each encompassing subsets of geometric primitives.  When performing queries the AABB tree quickly eliminates large regions of space by checking if the query point falls within the current bounding box. This hierarchical pruning dramatically reduces the number of tetrahedra that must be explicitly checked for containment, leading to substantial performance gains and enabling efficient, scalable querying of high-density meshes.

The implementation of the described methodology is encapsulated in the \texttt{SimulationField} node. Users must provide a \texttt{.inp} file defining the mesh, a CSV file containing nodal simulation results, and a set of math expressions mapping simulation variables (e.g., displacement magnitude, stress, temperature) to material volume fractions. The \texttt{SimulationField} node interprets these expressions as volumetric fraction functions dependent on simulation results. Variables accessible within these expressions include scalar components (e.g., \texttt{dx}, \texttt{dy}, \texttt{dz}) and the magnitude of vector quantities (\texttt{len}).

Listing \ref{lst:bike_seat_code} demonstrating how to create a multi-material design informed by simulation results using the \texttt{SimulationField} node provided by the new \texttt{pyvcad} module.

\begin{listing}[h]
    \caption{\texttt{pyvcad} Code Bike Seat Example}
    \label{lst:bike_seat_code}
    \begin{minted}{python}
import pyvcad as pv
materials = pv.default_materials()

inp_path = "bike_seat_new.inp"
point_map_path = "bike_seat_point_map.csv"
expressions = ["len-0.000055)/0.00035", 
               "-(len-0.000055)/0.00035+1"]
materials = [materials.id("blue"),
             materials.id("green")]

root = pv.SimulationField(inp_path, point_map_path, 
                          expressions, materials)
    \end{minted}
\end{listing}

\subsubsection{Case Study: Simulation-informed Bike Seat Design}
\label{sec:simulation_field}
\begin{figure}[h]
    \centering
    \includegraphics[width=\linewidth]{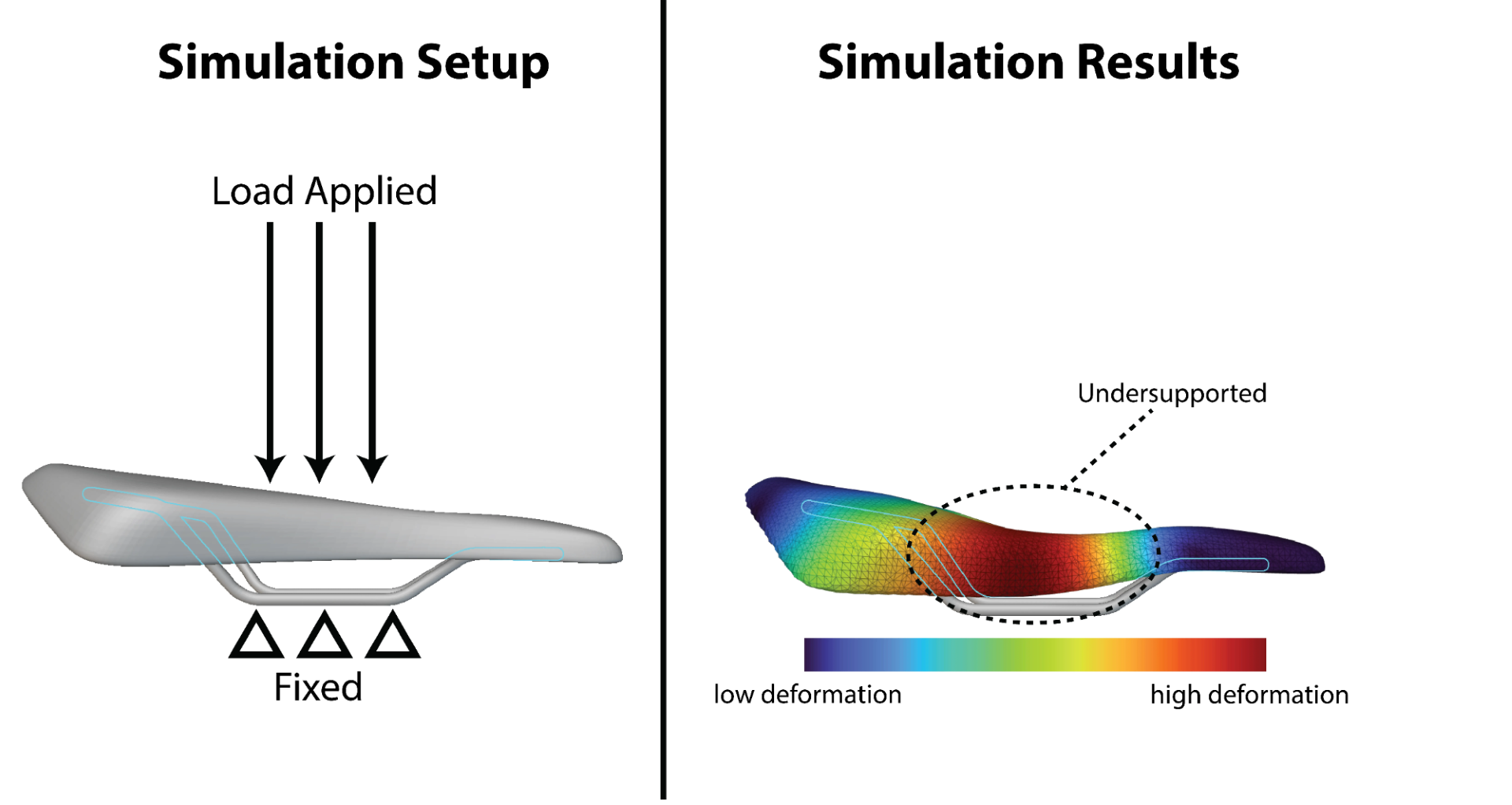}
    \caption{Simulation setup and corresponding results highlighting excessive deformation in the unsupported central region of a uniformly soft-material bike seat under rider load. Arrows indicate the downward force based on a rider sitting position.}
    \label{fig:seat-setup-results}
\end{figure}

To demonstrate the practical effectiveness of the \texttt{SimulationField} node, we apply it to a real-world design scenario: optimizing the material distribution within a 3D-printed bicycle seat. The primary goal in this case study is to mitigate excessive deformation, particularly in an intentionally unsupported central region, while simultaneously preserving rider comfort by controlling the material composition in the design volume. Figure~\ref{fig:seat-setup-results} depicts the bike seat design used in this analysis, which features fixed mounting rails at both ends but no central structural support, as is typical in a bicycle seat design. To investigate performance, we simulated the seat using a single homogeneous soft material with properties similar to Stratasys PolyJet material Agilus30. The finite element model, discretized into tetrahedral elements, was subjected to loads representative of a rider’s weight. From figure \ref{fig:seat-setup-results} we see the simulations, conducted using nTopology, revealed significant deformation concentrated in the unsupported central area, highlighting structural inadequacies when composed uniformly of the soft material. We chose to use nTopology in this example, differing from the previous examples using ABAQUS, to demonstrate how our method works with multiple FEA solvers. 

Ideally, the entire seat would be composed of the softer material to maximize comfort. However, given the structural demands identified by the simulation, it is clear that selective incorporation of a more rigid material is necessary to improve structural performance. A homogeneous mixture of soft and rigid materials across the entire seat could resolve deformation issues but would unnecessarily compromise rider comfort in adequately supported regions. Instead, we used the \texttt{SimulationField} node in OpenVCAD to dynamically vary the material mixture based on the simulated displacement magnitude. Specifically, the simulation results, provided as nodal displacements from the original tetrahedral mesh were exported from nTopology. nTopology was only used as the linear elastic simulator in this case study. The simulation results were imported into OpenVCAD, along with a set of user-defined mathematical expressions mapping displacement magnitudes to corresponding volume fractions of rigid (Vero-like) and soft (Agilus30-like) materials.

The Python snippet in Listing~\ref{lst:bike_seat_code} illustrates the process of defining the \texttt{SimulationField} node with the required parameters: the tetrahedral mesh (.inp file), nodal displacement data (CSV), and material fraction expressions. This method automatically interprets the simulation results, converting them into a spatially varying, functionally graded material distribution.

\begin{figure}[h]
    \centering
    \includegraphics[width=\linewidth]{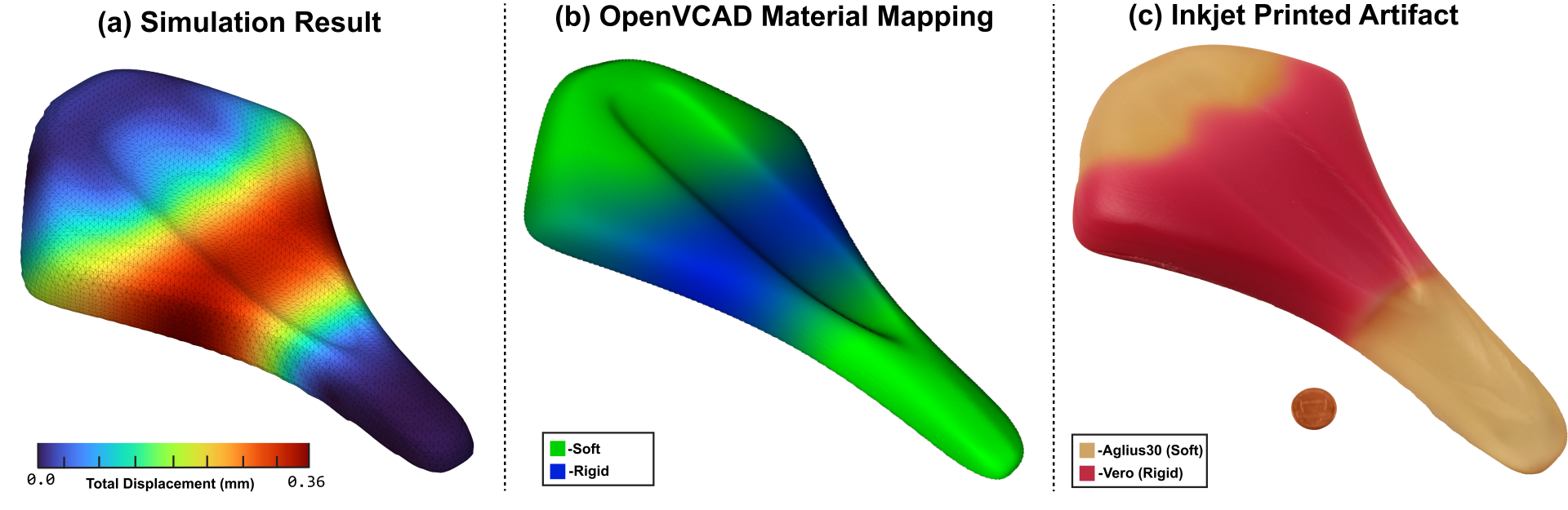}
    \caption{Bike seat case study results demonstrating simulation-informed multi-material optimization: (a) simulated displacement field, (b) OpenVCAD-generated multi-material distribution informed by displacement magnitude, and (c) final printed artifact fabricated using a Stratasys J750 printer with Vero (rigid) and Agilus30 (soft) photopolymer materials.}
    \label{fig:j750-bike-seat}
\end{figure}

The final multi-material design was exported from OpenVCAD for fabrication on a Stratasys J750 Inkjet 3D printer by sampling approximately 26.2 billion voxels, outputting the voxel data into PNG image stacks in 24 minutes. Figure~\ref{fig:j750-bike-seat} compares the initial simulation displacement field obtained from nTopology, the corresponding OpenVCAD-generated multi-material gradient map, and the fabricated artifact. As intended, the rigid material is localized within areas of significant deformation, while regions adequately supported by the rails remain predominantly composed of soft material. This approach successfully balances structural integrity and rider comfort, demonstrating the integration capabilities of OpenVCAD to optimize multi-material additive manufacturing designs informed directly by external simulation data.

\section{Exporting as Meshes}

To broaden compatibility with widely adopted slicing platforms, our work presents a novel compiler module capable of exporting volumetric, functionally graded designs into sets of standard triangulated mesh files. Although this approach forgoes the fine-grained process-parameter integration achievable through direct toolpath coupling outlined in Wade et al. 2025, it permits spatially varying material properties or slicing parameters to be approximated using standard slicing software \cite{wade_implicit_2025}. The mesh export method takes as input a user-specified number of output meshes, effectively segmenting the continuous material fraction space into discrete sub-regions. 

A widely recognized algorithm for generating a triangulated mesh from implicit signed-distance representations is the Marching Cubes algorithm~\cite{lorensen_marching_1987}. This algorithm identifies surfaces by sampling volumetric scalar fields at regular intervals, generating triangle approximations of the implicit surface. However, Marching Cubes alone only captures geometric boundaries and does not inherently account for multi-material gradients. To address this, we introduce a hybrid sampling and segmentation method described by Algorithm~\ref{alg:mesh-export}.

\begin{algorithm}[h]
\caption{Hybrid Segmentation Algorithm for Multi-material Mesh Export}
\label{alg:mesh-export}
\begin{algorithmic}[1]
\Require OpenVCAD design $D$, number of output meshes $N$, voxel resolution $r$
\Ensure Set of triangulated meshes $\{M_i\}_{i=1}^{N}$
\State Divide the material fraction space into $N$ ranges $\{R_i\}_{i=1}^{N}$
\State Sample design $D$ at resolution $r$ to produce a voxel grid $G$ storing signed distances
\ForAll{material ranges $R_i$}
    \State Initialize filtered voxel grid $G_i$
    \ForAll{voxels $v \in G$}
        \State Query $D$ at voxel location to get multi-material volume fractions
        \If{fractions fall within range $R_i$}
            \State Copy signed distance from $v$ to $G_i$
        \Else
            \State Mark voxel as exterior in $G_i$
        \EndIf
    \EndFor
    \State Apply Marching Cubes to $G_i$ to generate mesh $M_i$
\EndFor
\State \Return segmented meshes $\{M_i\}_{i=1}^{N}$
\end{algorithmic}
\end{algorithm}

\subsection{Case Study: Graded Slicing Settings for a Bike Seat}

\begin{figure*}[h]
    \centering
    \includegraphics[width=\linewidth]{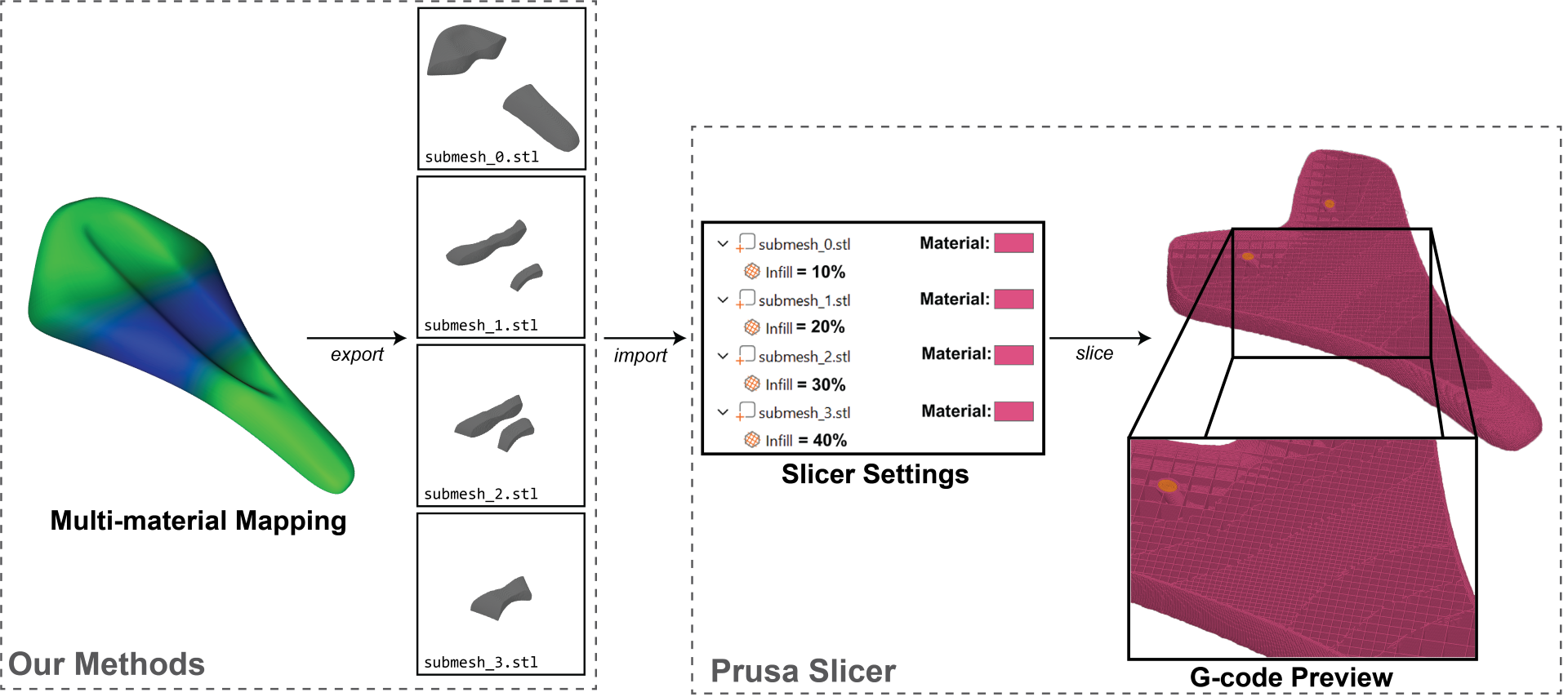}
    \caption{OpenVCAD bike seat segmented into four meshes based on material volume fractions. The corresponding g-code generated using Prusa Slicer is shown alongside the slicer settings, demonstrating spatially varying infill densities based on segmented meshes.}
    \label{fig:bike-seat-mesh-slicing}
\end{figure*}

To illustrate the utility of this mesh segmentation method, we revisit the previously discussed bike seat design optimized through simulation-informed multi-material distribution (Section~\ref{sec:simulation_field}). Figure~\ref{fig:bike-seat-mesh-slicing} shows the simulation-informed multi-material bike seat segmented into four discrete meshes based on volume fraction ranges of the two base materials. A sample resolution of $0.5\,\mathrm{mm}$ was used, resulting in approximately 20 million voxel queries and four exported meshes, completed in less than 30 seconds.

These segmented meshes were imported into Prusa Slicer, where each mesh was assigned distinct infill densities interactively. Despite using a single flexible TPU (90A) material for the entire print, the variable infill density replicated the rigidity gradient suggested by the original simulation. Specifically, regions containing higher proportions of rigid material in the OpenVCAD design were assigned greater infill densities. The resulting g-code toolpath with spatially varying infill densities is depicted in Figure~\ref{fig:bike-seat-mesh-slicing}(b). For smoother gradient transitions, a higher number of mesh segments can be exported and individually assigned slicing parameters.

\begin{figure}[h]
    \centering
    \includegraphics[width=\linewidth]{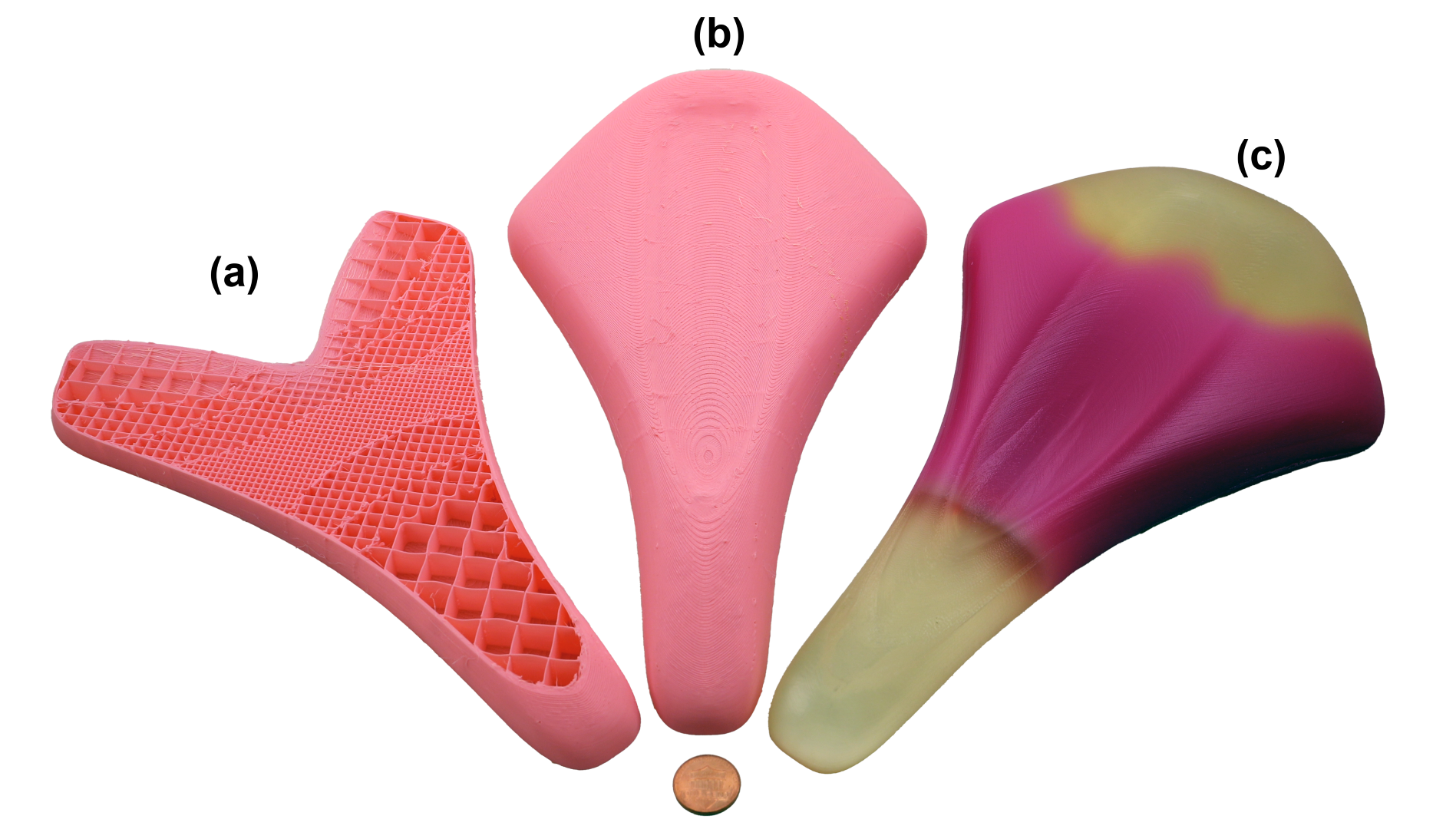}
    \caption{Fabricated bike seat artifacts demonstrating mesh-based export capability: (a) cross-sectional view showing spatial infill gradients, (b) complete seat printed with single TPU material using material extrusion, and (c) PolyJet print comparison using Vero (rigid) and Agilus30 (soft) materials.}
    \label{fig:bike-seat-printed-results}
\end{figure}

The printed artifact was fabricated on a Prusa XL material extrusion system using a single TPU filament. Figure~\ref{fig:bike-seat-printed-results} compares the material extrusion printed artifact (with cut-away revealing variable infill densities) against the PolyJet print presented earlier. This figure underscores the versatility of OpenVCAD in exporting the same underlying design to various additive manufacturing modalities through interchangeable export modules. Although the material gradient fidelity achievable via voxel-based Inkjet printing is higher, this mesh-based export significantly expands compatibility, enabling OpenVCAD designs to be realized using widely available extrusion-based systems.

\section{Conclusion}
In this work, we presented advancements for the open-source OpenVCAD framework. The introduction of a Python-based API (\texttt{pyvcad}) marks a fundamental shift from the previous domain-specific language, increasing parametric expressiveness through standard programming constructs and enables integration with external libraries. Additionally, we introduced novel implicit modeling techniques for multi-material lattice structures, leveraging primitives like the \texttt{GraphLattice} and \texttt{Tile} nodes. Our methods support comprehensive spatial grading strategies including global, per-cell, and per-strut, enabling detailed tuning of local material properties.

We demonstrated robust integration with finite element analysis workflows, including direct export to Abaqus compatible meshes with adaptive element sizing based on local material heterogeneity. Furthermore, we showed how simulation results can be imported back into OpenVCAD through the \texttt{SimulationField} node, enabling automated generation of spatially graded materials informed by mechanical responses. Our case studies, including graded bars, color calibration sheets, and an algorithmically generated multi-material Stanford Bunny lattice, highlight the design expressiveness and automation capabilities of our methods. Additionally, simulation-driven examples, such as functionally graded lattice structures and the multi-material bicycle seat optimized for comfort and structural performance, highlighted the benefits of integration with simulation software.

The enhanced capabilities introduce new challenges, notably the increased complexity in translating spatially graded material fractions into target physical properties. Designers must  possess knowledge linking volume fractions to performance characteristics such as shore hardness, density, and modulus, underscoring the need for an automated translation process. Additional future work could focus on deeper integration with advanced simulation methods, particularly exploring automated design synthesis with multi-material geometry generation and simulation in a closed loop with an optimizer. Further extensions could incorporate non-mechanical simulation domains like thermal and electrical performance, broadening the applicability of the framework. Future work could also explore integrating OpenVCAD directly with commercial slicing tools through plugins, enabling automatic assignment of materials and process settings to exported segmented meshes. A promising avenue is the use of the \texttt{3MF} format, which supports embedding XML-based metadata on a per-model basis \cite{3MF_spec}. Many slicers, such as PrusaSlicer, already leverage this functionality to specify per-model material assignments and print settings. Extending OpenVCAD to export segmented meshes with embedded configuration data inside a single \texttt{3MF} file would streamline multi-material workflows. Such capabilities would be particularly valuable for systems like the Bambu Labs AMS, which can print with up to 16 distinct materials within a single print \cite{BambuLab_ConnectAMS}.

The developments presented in this work lower the barrier to entry for designers, facilitating broader research and industrial adoption of fully volumetric, functionally graded design. By providing a Python-centric, open-source platform, we enable researchers to readily experiment and integrate advanced multi-material design into their additive manufacturing workflows. Towards this goal, we provide the \texttt{pyvcad} library on the PyPi package manager and include getting started documentation in the appendix. We invite practitioners, researchers, and collaborators to use, extend, and build upon these tools, fostering collective advancement toward advancements in computational design for multi-material additive manufacturing.

\begin{acks}
This material is based upon work supported by the Charles Stark Draper Laboratory, Inc. under Contract No. N00030-24-C-6001. Any opinions, findings and conclusions or recommendations expressed in this material are those of the author(s) and do not necessarily reflect the views of Strategic Systems Programs.
\end{acks}

\bibliographystyle{ACM-Reference-Format}
\bibliography{references}

\appendix
\section{Appendix: Getting Started with the \texttt{pyvcad} Library}
The Python bindings for OpenVCAD can be installed using the command \texttt{pip install OpenVCAD}. This installs three modules: (1) \texttt{pyvcad}, which includes the core OpenVCAD library for representing multi-material designs; (2) \texttt{pyvcadviz}, which provides a VTK-based renderer for \texttt{pyvcad} objects; and (3) \texttt{pyvcad\_compilers}, which contains reference compilers for exporting PNG stacks, surface meshes, and FEA-compatible meshes.

\begin{figure*}[h]
    \centering
    \includegraphics[width=\linewidth]{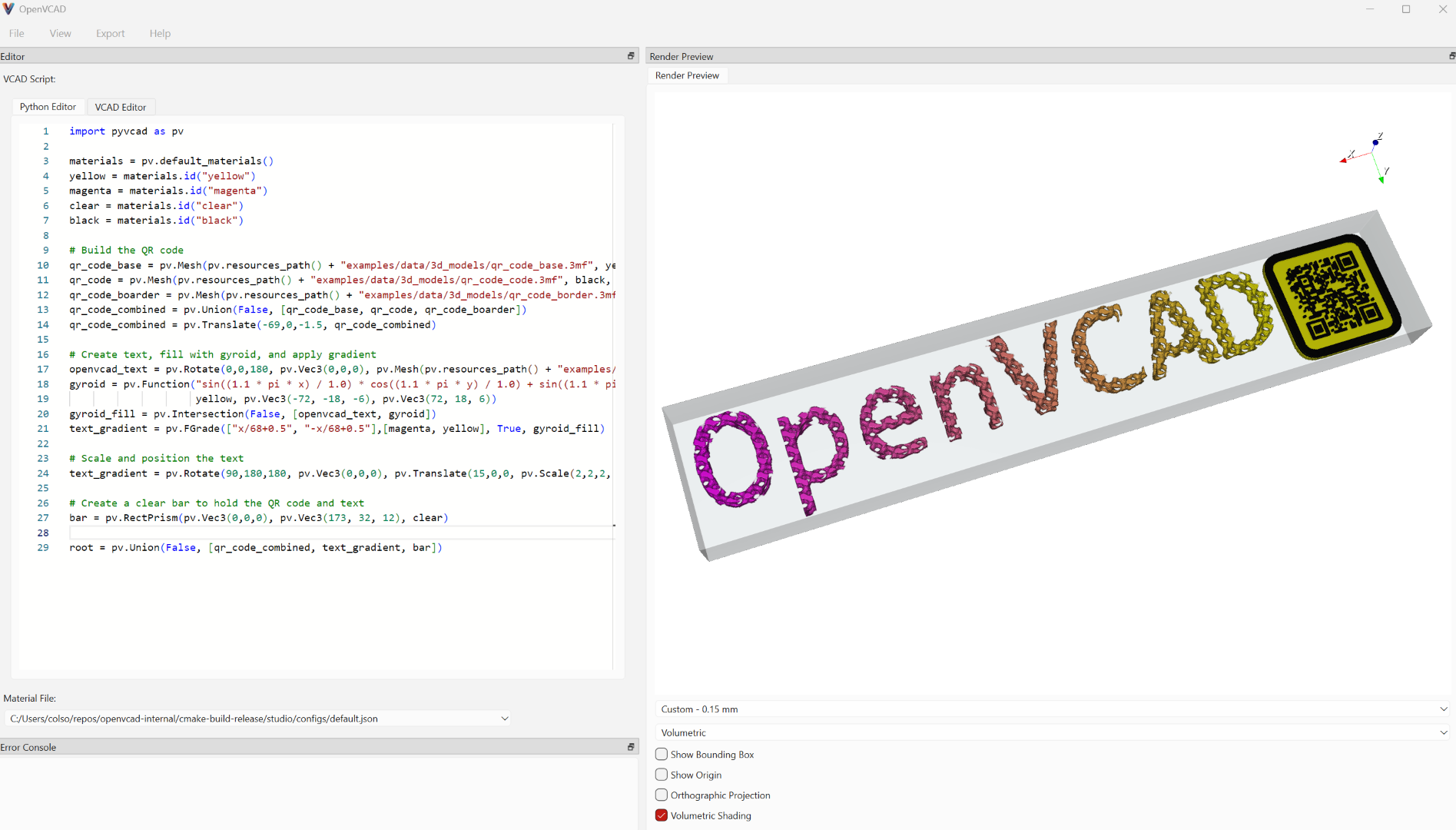}
    \caption{The OpenVCAD Studio Integrated Development Environment with an object loaded as code and rendered volumetrically.}
    \label{fig:vcad-studio}
\end{figure*}

An integrated development environment (IDE) that includes a text editor, renderer, and compiler interface in a single UI, called VCAD Studio, is shown in Figure \ref{fig:vcad-studio}. This program is available for download for Windows and MacOS on the \href{https://github.com/MacCurdyLab/OpenVCAD-Public}{OpenVCAD GitHub page}. All example code and data used in this paper are provided in the supplementary materials.

\end{document}